\documentclass[11pt]{article}
\usepackage{./rmylay,flushend}
\usepackage{xspace}

\newcommand{\cU}{{\cal U}} 

\renewcommand{\eg}{{\emph{e.g.\ }}}
\newcommand{\iid}{\emph{i.i.d.}\xspace}

\newcommand{\ptr}{{\psi}}

\newcommand{\Ptr}{{\Psi}}

\newcommand{\x}{{\bf x}}

\newcommand{\X}{\cX}

\title{On redundancy of memoryless sources over countable alphabets}
\author{Maryam Hosseini
\and
Narayana Santhanam }

\begin{document}
\maketitle 
\begin{abstract}
  The minimum average number of bits need to describe a random variable is its
  entropy. This supposes knowledge of the distribution of the random
  variable. On the other hand, \emph{universal compression} supposes that the
  distribution of the random variable, while unknown, belongs to a known set
  $\cP$ of distributions.  Such \emph{universal descriptions} for the random
  variable are agnostic to the identity of the distribution in
  $\cP$. But because they are not matched exactly to the underlying distribution
  of the random variable, the average number of bits they use is higher, and 
  the excess over the entropy used is the \emph{redundancy}. This formulation 
  is fundamental to problems not just in compression, but also estimation and 
  prediction and has a wide variety of applications from language modeling to
  insurance.

  In this paper, we study the redundancy of universal encodings of strings
  generated by independent identically distributed (\iid) sampling from a set $\cP$ of
  distributions over a countable support. We first show that if describing a
  single sample from $\cP$ incurs finite redundancy, then $\cP$ is tight but
  that the converse does not always hold.  

  If a single sample can be described with finite \emph{worst-case-regret} (a more
  stringent formulation than redundancy above), then it is known that describing
  length-$n$ \iid samples only incurs a diminishing (in $n$) redundancy \emph{per symbol}
  as $n$ increases. However, we show it is possible that a collection $\cP$ incurs
  finite redundancy, yet description of length-$n$ \iid samples incurs a constant
  redundancy per symbol encoded. We then show a sufficient condition on $\cP$ such
  that length-$n$ \iid samples will incur diminishing redundancy per symbol
  encoded.
\end{abstract}

A number of statistical inference problems of significant contemporary interest, 
such as text classification, language modeling, and DNA microarray analysis, 
requires computing inferences based on observed 
sequences of symbols in which the sequence length or sample size
is comparable or even smaller than the set of symbols, the \emph{alphabet}.
For instance, language models for speech recognition estimate
distributions over English words using text examples much smaller than
the vocabulary. 

To model these problems, several lines of work have considered universal
compression over large alphabets. Generally, the idea here is to model the
problem at hand with a class of models $\cP$ instead of a single
distribution. The model underlying the data is assumed or known to belong to the
class $\cP$, but the exact identity of the model remains unknown. Instead,
we aim to use a universal description of data. 

The universal description uses more bits on an average (over the sample) than if
the underlying model was known, and the additional number of bits used by the
universal description is called the \emph{redundancy} against the true
model. The average excess bits over the entropy of the model will be refered to
as the \emph{model redundancy} for that model. Since one does not know the true
model in general, a common approach is to consider \emph{strong redundancy},
which is the supremum over all models of the class of the model redundancy.

Typically we look at sequences of \iid symbols, and therefore we usually 
refer to the redundancy of distributions over length-$n$ sequences obtained
by \iid sampling from distributions from $\cP$. The length $n$ of sequences
considered will typically be refered to as the sample size.

The nuances of prediction, compression or estimation where the alphabet size and
sample size are roughly equal are not well captured by the asymptotics of the
strong redundancy with of a class over a fixed alphabet and sample size going to
infinity. Rather, they are better captured when we begin with a countably
infinite support and let the sample size approach infinity, or when we let the
alphabet size scale as a function of the sample size.

To begin with, the collection of all \iid distributions over countably infinite
supports or alll \iid distributions over an alphabet whose size is comparable to
the sample length of interest have very high redundancy that renders most
estimation or prediction problems impossible. Therefore, there are several 
alternative formulations to tackle language modeling, or classification and
estimation questions over large alphabets.

\paragraph{Patterns} One line of work is the patterns~\cite{OSZ03} approach that considers the
compression of the \emph{pattern} of a sequence rather than the sequence itself.
Patterns abstract the identities of symbols, and indicate only the relative
order of appearance. For example, the pattern of TATTLE is 121134, while that of
HONOLULU is 12324545. The point to note is that patterns of length-$n$ \iid
sequences can be compressed (no matter what the underlying countably infinite
alphabet is) with redundancy that grows sublinearly in $n$, therefore the
excess bits needed is asymptotically diminishing (in $n$) per-symbol redundancy.
Indeed insights learnt in this line of work will be used to understand compression
of sequences as well in this paper.

\paragraph{Envelope on model classes} A second line of work considers restricted
model classes for applications, particularly where the collection of models can
be described in terms of an envelope~\cite{BGG08}. This approach leads to an
understanding of the worst case formulations. In particular, we are interested
in the result that if the \emph{worst-case} redundancy (different from, and a
more stringent formulation than the strong redundancy described here) of describing
a single sample is finite, then the per-symbol strong redundancy diminishes to 0. We
will interpret this result towards the end of the introduction.

\paragraph{Data derived consistency} A third line of work ignores the uniform
convergence framework underlying strong redundancy formulations. This is useful
for large or infinite alphabet model classes which have poor or no strong
redundancy guarantees, but ask a question that cannot be answered with the
patterns approach above.  In this line of work, one obtains results on the
\emph{model redundancy} described above instead of \emph{strong} redundancy
. For example, a model class is said to be weakly compressible if there is a
universal measure that ensures that for all models, the model redundancy
normalized by the sample size (per-symbol) diminishes to 0. The rate at which
the per-symbol model redundancy diminishes to 0 depends on the underlying model,
and for some models could be arbitrarily slower than others. Given a particular
blocklength, $n$, there may be hence no non-trivial guarantee that holds over
the entire model class unlike the strong redundancy formulation. 

But if we add on the additional constraint that we should be estimate the rate
of convergence from the data, we get the \emph{data-derived} consistency
formulations in~\cite{SAKS14}. Fundamental to further research in this direction
is a better understanding of how single letter redundancy (of $\cP$) relates to
the redundancy of length-$n$ strings (that of $\cP^n$). The primary theme of
this paper is to collect such results on strong redundancy of classes over
countably infinite support.

In the fixed alphabet setting, this connection is well understood. If the
alphabet has size $k$, the redundancy of $\cP$ is easily seen to be always
finite (in fact $\le \log k$) and that of $\cP^n$ scales as $\frac k2 \log
n$. But when $\cP$ does not have a finite support, the above bounds are
meaningless.

On the other hand, the redundancy of a class $\cP$ over $\naturals$ may in
general be infinite. But what about the case where redundancy of $\cP$ is
finite?  Now a well known \emph{redundancy-capacity}~\cite{MF98} argument can be
used to interpret the redundancy---and this tells us that the redundancy is the
amount of information we can get about the source from the data. In this case,
finite (infinite respectively) redundancy of $\cP$ implies that a single symbol
contains finite (infinite respectively) amount of information about the model.

The natural question then is does it imply that the redundancy of length-$n$
\iid strings from $\cP$ grows sublinearly? Equivalently, do finite redundancy
classes over $\naturals$ behave essentially like some of their fixed alphabet
counterparts?  In some formulations (the worst-case), this indeed holds.
Roughly speaking, this informs us that as the universal encoder sees more and
more of the sequence, it learns less and less of the underlying model. This 
would be in line with our intuition where seeing more data fixes the model,
so as we so more data there is less to learn.

To understand these connections, we first show that if the redundancy of a
collection $\cP$ of distributions over $\naturals$ is finite, then $\cP$ is
tight. This turns out to be a useful tool to check if the redundancy is finite
in~\cite{SAKS14} for example. 

But in a departure from previous formulations, we then demonstrate that it is
possible for a class $\cP$ to have finite redundancy, yet the redundancy of
length-$n$ strings sampled \iid from $\cP$ does not grow sublinearly in $n$!
Therefore, roughly speaking, no matter how much of the sequence the universal
encoder has seen, it learns at least a constant number of bits about the
underlying model each time it sees another symbol. No matter how much data we
see, there is more to learn! We finally obtain a sufficient condition on a class
$\cP$ such that the asymptotic per-symbol redundancy of length-$n$ \iid strings
diminishes to 0.

\ignore{To bring out the possible properties that go into ensuring that
  length-$n$ \iid strings have diminishing per-symbol redundancy, we consider
  several subclasses of monotone distributions such that the asymptotic
  per-symbol redundancy diminishes to 0.}

\newcommand{\cJ}{\cal J}
\section{Notation and background}
We will review the notions of universal compression, redundancy and patterns 
here, as well as some allied results that we will make use of in this paper.

Let $\cP$ be a collection of distributions over $\naturals$.  Let $\cP^n$ be the
set of distributions over length-$n$ sequences obtained by \iid sampling from
$\cP$. Let $\cP^\infty$ be the collection of measures over infinite length
sequences\footnote{Observe that $\cN^n$ is countable for every $n$. For
  simplicity of exposition, we will think of each length-$n$ string $\x$ as a
  subset of $\cN^\infty$---the set of all semi-infinite strings of naturals that
  begin with $\x$. Each subset of $\cN^n$ is therefore a subset of $\cN^\infty$.

  Now the collection $\cJ$ of all subsets of $\cN^n$, $n\ge 1$, is a
  semi-algebra~\cite{JR}.  The probabilities \iid sampling assigns to finite
  unions of disjoint sets in $\cJ$ is the sum of that assigned to the components
  of the union. Therefore, there is a sigma-algebra over the uncountable set
  $\cN^\infty$ that extends $\cJ$ and matches the probabilities assigned to sets
  in $\cJ$ by \iid sampling. The reader can assume the sigma-algebra is the
  minimal such extension. $\cP^\infty$ is the measure on this sigma-algebra that
  matches what the probabilities \iid sampling gives to sets in $\cJ$. See,
  \eg~\cite{JR}, for a development of elementary measure theory that lays out
  the above results.} of $\naturals$ obtained by \iid sampling from
distributions of $\cP$.

\subsection{Redundancy}
\subsubsection{Strong compression} 

A class $\cP^\infty$ of measures over infinite sequences of
natural numbers is called strongly compressible if there is a measure $q$ over
$\naturals^\infty$ satisfying
\begin{equation}
\label{eq:str_rdn}
\limsup_{n\to\infty} 
\sup_{p\in\cP^\infty}
\frac1n E_p \log\frac{p(X^n)}{q(X^n)} =0.
\end{equation}
In particular, we call
\[
\inf_q \sup_{p\in\cP^\infty}
\frac1n E_p \log\frac{p(X^n)}{q(X^n)} 
\]
the redundancy of length-$n$ sequences, or length-$n$ redundancy. The single
letter redundancy refers to the special case when $n=1$. 

We will often be concerned with sequences of symbols drawn \iid from $\cP$, and
let $\cP^\infty$ be the measures induced on infinite sequences of naturals
obtained by \iid sampling from distributions in $\cP$. Our primary goal is to
understand the connections between the single letter redundancy on the one hand
and the behavior of length-$n$ \iid redundancy on the other.  Strong compression
length-$n$ redundancy can be seen as the capacity of a channel from $\cP$ to
$\naturals^n$, where the conditional probability distribution over $\naturals^n$
given $p\in\cP$ is simply the distribution $p$ over length-$n$ sequences.

We note that it is possible to define an even more stringent notion---a
\emph{worst case} formulation. For length-$n$ sequences, this is 
\[
\inf_q
\sup_{p\in\cP^\infty}
\frac1n \sup \log\frac{p(X^n)}{q(X^n)}.
\]
We will not concern ourselves with the worst case formulation in this paper, but
mention it in passing for comparisons.  In the worst case setting, finite single
letter redundancy is necessary and sufficient for the asymptotic per-symbol
worst case redundancy to diminish to 0. 

But we show in this paper that it is not necessarily the case for strong
redundancy. It is quite possible that classes with finite single letter strong
redundancy have asymptotic per-symbol strong redundancy bounded away from 0.

\subsection{Patterns}
\label{s:b:ptr}
Recent work~\cite{OSZ03} has formalized a similar framework for countably
infinite alphabets.  This framework is based on the notion of \emph{patterns} of
sequences that abstract the identities of symbols, and indicate only the
relative order of appearance. For example, the pattern of PATTERN is
1233456. The $k'$th distinct symbol of a string is given an index $k$ when it
first appears, and that index is used every time the symbol appears
henceforth. The crux of the patterns approach is to consider the set of measures
induced over patterns of the sequences instead of considering the set of
measures $\cP$ over infinite sequences,

Denote the pattern of a string $\x$ by $\Ptr(\x)$. There is only one possible
pattern of strings of length 1 (no matter what the alphabet, the pattern of a
length-1 string is 1), two possible patterns of strings of length 2 (11 and 12),
and so on. The number of possible patterns of length $n$ is the $n'$th Bell
number~\cite{OSZ03} and we denote the set of all possible length $n$ patterns
by $\Ptr^n$. The measures induced on patterns by a corresponding
measure $p$ on infinite sequences of natural numbers assigns to any pattern
$\ptr$ a probability
\[
p(\ptr)=p\Paren{\sets{\x : \Ptr(\x)=\ptr}}.
\]
In~\cite{OSZ03} the length-$n$ pattern redundancy,
\[
\inf_q 
\sup_{p\in\cP^\infty}
\frac1n E_p \log\frac{p(\Ptr(X^n))}{q(\Ptr(X^n))},
\]
\newcommand{\qptr}{q_{{}_\Psi}}
was shown to be upper bounded by $\pi (\log e)\sqrt{\frac{2n}3}$.
It was also shown in~\cite{S06:thesis} that there is a measure
$q$ over infinite length sequences which satisfies for all $n$ simultaneously
\[
\sup_{p\in\cP^\infty}
\frac1n E_p \log\frac{p(\Ptr(X^n))}{q(\Ptr(X^n))}
\le 
\pi (\log e)\sqrt{\frac{2n}3}
+\log (n(n+1)).
\]
Let the measure induced on patterns by $q$ be denoted as $\qptr$ for
convenience. 

We can (naively) interpret the probability estimator $\qptr$ as a sequential
prediction procedure that estimates the probability that the symbol $X_{n+1}$
will be ``new'' (has not appeared in $X_1^n$), and the probability that
$X_{n+1}$ takes a value that has been seen so far. This view of estimation also
appears in the statistical literature on Bayesian nonparametrics that focuses on
exchangeability.  Kingman \cite{Kin80} advocated the use of {\it exchangeable
  random partitions} to accommodate the analysis of data from an alphabet that
is not bounded or known in advance.  A more detailed discussion of the history
and philosophy of this problem can be found in the works of Zabell
\cite{Zab92,Zab97} collected in \cite{Zab05}.  

\subsection{Cummulative distributions and tight classes}
For our purposes, the cumulative distribution function of any probability distribution
$p$ on $\naturals$ is a function
$F_p ~:~ \reals^+ \cup \{\infty\} \to [0,1]$ defined in the following (slightly unconventional) way.
We obtain $F_p$ by first defining $F_p$ on points in the support of
$p$ in the way cumulative distribution functions are normally defined.
We define $F_p$ for all other 
nonnegative real numbers
by linearly interpolating between the values in the support of
$p$. Finally, $F_p(\infty) := 1$. 

Let $F_p^{-1} ~:~ [0,1] \mapsto \reals^+ \cup \{\infty\}$ denote the 
inverse function of $F_p$. Then $F_p^{-1}(x) =0$ for all $0\le x< F_p(0)$.
If $p$ has infinite support then $F_p^{-1}(1) = \infty$, else
$F^{-1}_p(1)$ is the smallest natural number $y$ such that
$F_p(y)=1$.

A class $\cP$ of distributions on $\naturals$ is defined to be \emph{tight} if 
for all $\gamma>0$, 
\[
\sup_{p\in\cP} F_p^{-1}(1-\gamma) <\infty.
\]


\section{Redundancy and tightness}
We focus on the single letter redundancy in this section, and explore the connections
between the single letter redundancy of a class $\cP$ and the tightness of $\cP$.

\bLemma
\label{lm:bndprc}
A class $\cP$ with bounded strong redundancy is tight. Namely, if the strong
redundancy of $\cP$ is finite, then for any $\gamma>0$
\[
\sup_{p\in\cP} F_p^{-1}(1-\gamma) <\infty.
\]
\Proof 
$\cP$ has
bounded strong redundancy. Let $q$ be a distribution over 
$\naturals$ such that
\[
\sup_{p\in\cP} D(p||q) <\infty,
\]
and we define $R=\sup_{p\in\cP} D(p||q)$.
It follows that for all $p\in\cP$ and any $m$, 
\[
p( \left|\log \frac{p(X)}{q(X)} \right| > m ) \le (R+(2\log e)/e)/m,
\]
To see the above, note that if $S$ is the set of all numbers such 
that $p(x)<q(x)$, a well-known convexity argument shows that
\[
\sum_x p(x)\log \frac{p(x)}{q(x)}
\ge
p(S)\log \frac{p(S)}{q(S)}
\ge 
- \frac{\log e}e.
\]
We prove the lemma by contradiction. Pick $m$ so large that $(R+(2\log e)/e)/m<\gamma/2$.
For all $p$, we show that 
\[
p\Paren{ x: x\ge F_q^{-1}(1-\gamma/2^{m+1}) }\le \gamma.
\]
To see the above, observe that we can split the tail $x\ge F_q^{-1}(1-\gamma/2^{m+1})$
into two parts---(i) numbers $x$ such that
$\log \frac{p(x)}{q(x)} > m$. This set has probability $<\gamma/2$ under $p$.
(ii) remaining numbers $x$ such that
$\log \frac{p(x)}{q(x)} < m$. This set has probability $\le \gamma/2^{m+1}$ under
$q$, and therefore probability $\le \gamma/2$ under $p$.
The lemma follows.
\eLemma

The converse is not necessarily true. Tight classes need not have finite single
letter redundancy as the following example demonstrates.

\paragraph{Construction} Consider the following class $\cI$ of distributions over
$\naturals$.  First partition the set of natural numbers into the sets $T_i$,
$i\ge0$, where
\[
T_i = \sets{2^k\upto 2^{k+1}-1}.
\]
Note that $|T_k|=2^k$.
Now, $\cI$ is the collection of all possible distributions that can be formed as
follows. For all $i\ge 1$, we pick exactly one element of $T_i$ and assign it
probability $1/( i(i+1))$. Note that the set $\cI$ is uncountably infinite.\hfill$\Box$

\bCorollary
The set $\cI$ of distributions is tight. 
\Proof For all $p\in\cI$,
\[
\sum_{n\ge 2^k} p(n) = \frac1{k+1},
\]
namely, all tails are uniformly bounded over the class $\cI$ to ensure
insurability. Put another way, for all $\delta>0$ and all distributions
$p\in\cI$,
\[
F_p^{-1}(1-\delta) \le 2^{\floor{\frac1\delta}}-1.\eqed
\]
\eCorollaryp

On the other hand,
\bProposition
The collection $\cI$ does not have finite redundancy. 
\Proof
Suppose $q$ is any distribution over $\naturals$. We will show that 
$\exists p \in \cI$ such that 
\[
\sum_{n\ge 1} p(n) \log \frac{p(n)}{q(n)} 
\]
is not finite. Since the entropy of every $p\in\cI$ is finite, we just have to
show that for any distribution $q$ over $\naturals$, there $\exists p \in \cI$
such that
\[
\sum_{n\ge 1} p(n) \log \frac{1}{q(n)}
\]
is not finite.

Consider any distribution $q$ over $\naturals$. Observe that for all
$i$, $|T_i|=2^i$. It follows that for all $i$ there is $x_i\in T_i$ such that
\[
q(x_i) \le \frac1{2^i}.
\]
But by construction, $\cI$ contains a distribution $p$ that has for its support
$\sets{x_i:i\ge1}$ identified above. Furthermore $p$ assigns
\[
p(x_i)=\frac1{i(i+1)} \qquad\forall \,i\ge1.
\]
The KL divergence from $p$ to $q$ is not finite and the Lemma follows.  \eProposition

\section{Length-$n$ redundancy}
We study how the single letter properties of a collection $\cP$ of distributions
influences the compression of length-$n$ strings obtained by \iid sampling from distributions 
in $\cP$. Namely, we try to characterize when the length-$n$ redundancy of $\cP^\infty$
grows sublinearly in the blocklength $n$.

\bLemma 
\label{lm:rlb}
Let $\cP$ be a class of distributions over a countable support $\cX$.
For some $m\ge1$, consider $m$ pairwise disjoint subsets $S_i\subset \cX$
($1\le i \le m$) and let $\delta>1/2$. If there exist $p_1\upto p_m\in\cP$ such
that
\[
p_i(S_i)\ge \delta,
\]
then for all distributions $q$ over $\X$, 
\[
\sup_{p\in\cP} D(p||q) \ge \delta\log m.
\]
In particular if there are an infinite number of sets $S_i$, $i\ge 1$ and
distributions $p_i\in\cP$ such that $p_i(S_i)\ge \delta$, then the redundancy is
infinite.  \Proof This is a simplified formulation of the
\emph{distinguishability} concept in~\cite{MF98}.  For a proof, see \eg~\cite{OS14}.
\eLemma

\subsection{Counterexample}
\label{s:ce}
We now show that it is possible for the single letter redundancy of a class
$\cB$ of distributiosn to be finite, yet the asymptotic per-symbol redundancy
(the length-$n$ redundancy of $\cB^\infty$ normalized by $n$) remains bounded
away from 0 in the limit the blocklength goes to infinity. To show this, we 
obtain such a class $\cB$.

\paragraph{Construction} As before partition the set $\naturals$ into
$T_i=\sets{ 2^i\upto 2^{i+1}-1}$, $i\ge 0$.  Recall that $T_i$ has $2^i$
elements. For all $\epsilon>0$, let $n_\epsilon=\floor{\frac1\epsilon}$.  Let
$1\le j\le 2^{n_\epsilon}$ and let $p_{\epsilon,j}$ be a distribution on
$\naturals$ that assigns probability $1-\epsilon$ to the number 1 (or
equivalently, to the set $T_0$), and $\epsilon$ to the $j'$th smallest element
of $T_{n_{\epsilon}}$, namely the number $2^{n_\epsilon}+j-1$. $\cB$ (mnemonic
for binary, since every distribution has at support of size 2) is the collection
of distributions $p_{\epsilon,j}$ for all $\epsilon>0$ and $1\le j \le
2^{n_\epsilon}$.  $\cB^\infty$ is the set of measures over infinite sequences of
numbers corresponding to \iid sampling from $\cB$.\hfill$\Box$

\bProposition For all $n\ge1$,
\[
\inf_q \sup_{p\in B^\infty}  E_p \log \frac{p(X^n)}{q(X^n)} \ge n\Paren{1-\frac1n}^n.
\]
\Proof
For all $n$, define $2^n$ pairwise disjoint sets $S_i$ of $\naturals^n$, $1\le i\le 2^n$, where
\[
S_i = \sets{1, 2^n+i-1}^n -\sets{1^n}
\]
is the set of all length-$n$ strings containing at most two numbers ($1$ and
$2^n+i-1$) and at least one occurance of $2^n+i-1$. Clearly, for distinct $i$
and $j$ between 1 and $2^n$.  $S_i$ and $S_j$ are disjoint.  Furthermore, the
distribution $p_{\frac1n,i}$ assigns $S_i$ the probability
\[
p_{\frac1n,i}(S_i) = 1-\Paren{1-\frac1n}^n > 1-\frac1e.
\]
From Lemma~\ref{lm:rlb}, it follows that length-$n$ redundancy of $\cB^\infty$ 
is lower bounded by
\[
\Paren{1-\frac1e}\log 2^n
= n\Paren{1-\frac1e}.\eqed
\]
\ePropositionp

\subsection{Sufficient condition}
In this section, we show a sufficient condition on single letter marginals
of $\cP$ and its redundancy that allows for \iid length-$n$ redundancy
of $\cP^\infty$ to grow sublinearly with $n$. This condition is, however,
not necessary---and the characterization of a condition that is both
necessary and sufficient is as yet open.

For all $\epsilon>0$, let $A_{p,\epsilon}$ is the set of all elements in the
support of $p$ which have probability $\ge \epsilon$, and let $T_{p,\epsilon}=\naturals-A_{p,\epsilon}$ 
For all $i$, the sets
\[
G_i =\sets{ x^i : A_{p,\frac{2\log i}i} \subseteq \sets{x_1,x_2\upto x_i}}
\]
where in a minor abuse of notation, we use $\sets{x_1\upto x_i}$ to denote the
set of distinct symbols in the string $x_1^i$. Let $B_i = \naturals - G_i$. Observe from
an argument similar to the coupon collector problem~\cite{} that (correct for
log 2 bases)
\bLemma
\label{lm:pb}
For all $i\ge1$,
\[
p(B_i)\le \frac{j}{2\log j} \Paren{1-\frac{2\log j}j}^j\le \frac1{j\log j}.\eqed
\]
\eLemmap

\bTheorem 
\label{thm:sff}
Suppose $\cP$ is a class of distributions over $\naturals$.  Let the
entropy of $p\in\cP$, denoted by $H(p)$, be uniformly bounded over the entire
class, and in addition let the redundancy of the class be finite. Namely,
\[
\sup_{p\in\cP} \sum_{x\in\naturals} p(x)\log\frac1{p(x)} <\infty
\quad\text{  and  }\exists q_1\text{ over $\naturals$ s.t. }\quad
\lim_{\delta\to0}
\sup_{p\in\cP} \sum_{x\in\naturals} p(x)\log\frac{p(x)}{q_1(x)} <\infty.
\]
Recall that for any distribution $p$, the set $T_{p,\delta}$ denotes the support
of $p$ all of whose probabilities are $<\delta$. Let 
\begin{equation}
\label{eq:cnd}
\lim_{\delta\to0}
\,\sup_{p\in\cP} \sum_{x\in T_{p,\delta}} p(x)\log\frac1{p(x)} =0
\quad\text{  and  }\exists q_1\text{ over $\naturals$ s.t. }\quad
\lim_{\delta\to0}
\sup_{p\in\cP} \sum_{x\in T_{p,\delta}} p(x)\log\frac{p(x)}{q_1(x)} =0.
\end{equation}
Then, the redundancy of length-$n$ distributions obtained by \iid sampling from distributions
in $\cP$, denoted by $R_n(\cP^\infty)$, grows sublinearly
\[
\limsup_{n\to \infty} \frac1n R_n(\cP^\infty) = 0.
\]
\Proof Let $\qptr$ be the optimal universal pattern encoder over patterns of
\iid sequences from Section~\ref{s:b:ptr}. Since the redundancy of $\cP$ is
finite, let $q_1$ be a universal distribution over $\naturals$ that attains
finite redundancy for $\cP$.  We consider a universal encoder as follows:
\begin{align*}
q(x^n) &= 
q(x^n,\Ptr(x^n)) \\
&\aeq{(a)}
q(\psi_1,x_1,\psi_2,x_2\upto \psi_n,x_n)\\
&=
\prod_{i\ge1} q(\ptr_i|\ptr^{i-1}_1,x^{i-1}_1) \prod_{j\ge1} q(x_j|\ptr^{j}_1, x^{j-1}_1)\\
&\ed
\prod_{i\ge1} q_\Psi(\ptr_i|\ptr^{i-1}_1) \prod_{j\ge1} q(x_j|\ptr^{j}_1, x^{j-1}_1)
\end{align*}
where in $(a)$, we denote $\Ptr(x^n)=\ptr_1\upto \ptr_n$. Furthermore we define
for all $x_1^{i-1}\in\naturals^{i-1}$ and all $\ptr^{i}\in\Ptr^i$ such that
$\ptr^{i-1}=\Ptr(x^{i-1})$,
\[
q(x_i|\ptr^{i}_1, x_1^{i-1})
\ed
\begin{cases}
1 & \text{if } x_i\in\sets{x_1\upto x_{i-1}} \text{ and } \Ptr(x^i)=\ptr^i\\
q_1(x_i) & \text{if } x_i\notin\sets{x_1\upto x_{i-1}} \text{ and } \Ptr(x^i)=\ptr^i.
\end{cases}
\]
Namely, we use an optimal universal pattern encoder over patterns of \iid
sequences, and encode any new symbol using a universal distribution over
$\cP$.
We now bound the redundancy of $q$ as defined above. We have for all $p\in\cP^\infty$,
\begin{align*}
E_p\log \frac{p(X^n)}{q(X^n)}
&=
\sum_{x^n} p(x^n)
\log 
\prod_{i\ge1} \frac{p(\ptr_i|\ptr^{i-1}_1,x^{i-1}_1) }{ q_\Psi(\ptr_i|\ptr^{i-1}_1)}
\prod_{j\ge1} \frac{p(x_j|\ptr^{j}_1, x^{j-1}_1)}{ q(x_j|\ptr^{j}_1, x^{j-1}_1)}\\
&=
\sum_{x^n} 
p(x^n)
\sum_{i=1}^n
\log 
\frac{
p(\ptr_i|\ptr^{i-1},x^{i-1}_1) }
{q_\Psi(\ptr_i|\ptr^{i-1}) }
+
\sum_{x^n} p(x^n)
\sum_{j=1}^n
\log 
\frac
{p(x_j|\ptr^{j}_1, x^{j-1}_1)}
{q(x_j|\ptr^{j}_1, x^{j-1}_1)}
\end{align*}
The first term, normalized by $n$, can be upper bounded by as follows
\begin{align*}
\frac1n
\sum_{x^n} 
p(x^n)
\sum_{i=1}^n
\log 
\frac{
p(\ptr_i|\ptr^{i-1},x^{i-1}_1) }
{q_\Psi(\ptr_i|\ptr^{i-1}) }
&\le
\frac1n\sum_{i=1}^n
\sum_{x^i} 
p(x^i)
\log 
\frac{p(\ptr_i|\ptr^{i-1},x^{i-1}_1) }
{p(\ptr_i|\ptr^{i-1}) }  + \pi\sqrt{\frac{2}{3n}}\\
&=
\frac1n\sum_{i=1}^n (H(\Ptr_i|\Ptr^{i-1})-H(\Ptr_i|X^{i-1}))+ \pi\sqrt{\frac{2}{3n}}.
\end{align*}
Now
\begin{align*}
H(\Ptr_i|\Ptr^{i-1})-H(\Ptr_i|X^{i-1})
&\le
H(X_i|\Ptr^{i-1})-H(\Ptr_i|X^{i-1})\\
&=
H-H(\Ptr_i|X^{i-1})\\
&= 
\sum_{x^{i-1}}p(x)
\!\!\!\!\!\!
\sum_{x\notin \sets{x_1\upto x^{i-1}}} 
\!\!\!\!\!\!
p(x)\log\frac1{p(x)}\\
&\le
p(G_{i-1}) 
\!\!\!
\sum_{x\in T_{p,2\frac{log(i-1)}{i-1}}}  
\!\!\!
p(x)\log \frac1{p(x)}
+
p(B_{i-1}) H\\
&\le
\sum_{x\in T_{p,2\frac{log(i-1)}{i-1}}}  
\!\!\!\!\!\!\!\!\!
p(x)\log \frac1{p(x)}
+
\frac{H}{(i-1)\log (i-1)}.
\end{align*}
We have split the length $j$ sequences into the sets $G_j$ and $B_j$ and use
separate bounds on each set that hold uniformly over the entire model class.
The last inequality $(a)$ above follows from Lemma~\ref{lm:pb}.
From condition~\eqref{eq:cnd} of the Theorem, we have that
\begin{flalign*}
&\lim_{i\to\infty} 
\sup_{p\in\cP}
\sum_{x\in T_{p,2\frac{log(i-1)}{i-1}}}  
\!\!\!\!\!\!\!\!\!
p(x)\log \frac1{p(x)}
=0\\
&\text{implying that }
\lim_{n\to\infty}
\sup_{p\in\cP}
\frac1n\sum_{i=1}^n 
\sum_{x\in T_{p,2\frac{log(i-1)}{i-1}}}  
\!\!\!\!\!\!\!\!\!
p(x)\log \frac1{p(x)}
+
\frac{H}{(i-1)\log (i-1)}
=0.
\end{flalign*}
For the second term, we have
\begin{align*}
\sum_{x^n} p(x^n)
\sum_{j=1}^n
\log 
\frac
{p(x_j|\ptr^{j}_1, x^{j-1}_1)}
{q(x_j|\ptr^{j}_1, x^{j-1}_1)}
&=
\sum_{j=1}^n
\sum_{x^j} p(x^j)
\log 
\frac
{p(x_j|\ptr^{j}_1, x^{j-1}_1)}
{q(x_j|\ptr^{j}_1, x^{j-1}_1)}\\
&\le
\sum_{j=1}^n
\sum_{x^{j-1}} p(x^{j-1})
\!\!\!\!
\sum_{x_j\notin \cA(x^{j-1})} 
\!\!\!\!
p(x_j)
\log 
\frac
{p(x_j)}
{q_1(x_j)}\\
&\le
\sum_{j=1}^n
\Paren{p(G_j)
\!\!\!\!
\sum_{x_j\notin A_{p,\frac{2\log j}j}}
\!\!\!\!
p(x_j)
\log 
\frac
{p(x_j)}
{q_1(x_j)}
+
Rp(B_j)}\\
&\le
\sum_{j=1}^n
\Paren{
\sum_{x_j\notin A_{p,\frac{2\log j}j}}
\!\!\!\!
p(x_j)
\log 
\frac
{p(x_j)}
{q_1(x_j)}
+
\frac{R}{j\log j}}.
\end{align*}
where as before, the last inequality is from Lemma~\ref{lm:pb}. Again from condition~\eqref{eq:cnd}, we have
\[
\sup_{p\in\cP}
\sum_{x_j\notin A_{p,\frac{2\log j}j}}
\!\!\!\!
p(x_j)
\log 
\frac
{p(x_j)}
{q_1(x_j)}
+
\frac{R}{j\log j}
=o(1)
\]
Therefore
\[
\sup_{p\in\cP}
\frac1n
\sum_{j=1}^n 
\Paren{
\sum_{x_j\notin A_{p,\frac{2\log j}j}}
\!\!\!\!\!\!
p(x_j)
\log 
\frac
{p(x_j)}
{q_1(x_j)}
+
\frac{R}{j\log j}}
=o(1)
\]
as well. The theorem follows.
\eTheorem

A few comments about~\eqref{eq:cnd} in Theorem~\ref{thm:sff} are in
order. Neither condition automatically implies the other. The set $\cB$ of
distributions in Section~\ref{s:ce} is an example where every distribution has
finite entropy, the redundancy of $\cB$ is finite,
\[
\lim_{\delta\to0}
\,\sup_{p\in\cB} \sum_{x\in T_{p,\delta}} p(x)\log\frac1{p(x)} =0
\quad\text{  but  }\forall q\text{ over $\naturals$ s.t. }\quad
\lim_{\delta\to0}
\sup_{p\in\cP} \sum_{x\in T_{p,\delta}} p(x)\log\frac{p(x)}{q_1(x)} >0.
\]
We will now construct another set $\cU$ of distributions over $\naturals$
such that every distribution in $\cU$ has finite entropy, the redundancy
of $\cU$ is finite, 
\begin{equation}
\label{eq:U}
\lim_{\delta\to0}
\,\sup_{p\in\cU} \sum_{x\in T_{p,\delta}} p(x)\log\frac1{p(x)} >0
\quad\text{  but  }\forall q\text{ over $\naturals$ s.t. }\quad
\lim_{\delta\to0}
\sup_{p\in\cP} \sum_{x\in T_{p,\delta}} p(x)\log\frac{p(x)}{q_1(x)} =0.
\end{equation}
At the same time, the length-$n$ redundancy of $\cU^\infty$ diminishes
sublinearly. This is therefore also an example to show that the conditions in
Theorem~\ref{thm:sff} are only sufficient, but in fact not necessary. It is yet
open to find a condition on single letter marginals that is both necessary and
sufficient for the asymptotic per-symbol redundancy to diminish to 0.

\paragraph{Construction} $\cU$ is a countable collection of distributions
$p_k$, $k\ge1$ where 
\[
p_k(x)
=
\begin{cases}
1-\frac1{k^2} & x=0,\\
\frac1{k^2 2^{k^2}} & 1\le x\le 2^{k^2}.
\end{cases}\eqed
\]
The entropy of $p_k\in\cU$ is therefore $1+h\Paren{\frac1{k^2}}$. Note that the
redundancy of $\cU$ is finite too. To see this, note that
\begin{equation}
\label{eq:wru}
\sum_{n \ge 1} 
\sup_{k\ge 1} 
p_k(n)
\le 
\sum_{n \ge 1} 
\sum_{p_k: k\ge 1} 
p_k(n)
=
\sum_{p_k: k\ge 1} 
\sum_{n \ge 1} 
p_k(n)
=
\sum_{p_k: k\ge 1} 
\frac1{k^2}
=\frac{\pi^2}6.
\end{equation}
Furthermore, letting 
$
R\ed
\log \Paren{\sum_{n \ge 1} 
\sup_{k\ge 1} 
p_k(n)}
$,
we have that the distribution
\[
q(n)
=
\begin{cases}
1/2 & n=0,\\
\frac{\sup_{k\ge1} p_k(n)}{2^{R+1}} & n\ge 1.
\end{cases}
\] 
satisfies for all $p_k\in\cU$
\[
\sum_{n\ge0} p_k(n) \log \frac{p_k(n)}{q(n)}
\le 
1+\frac{R+1}{k^2} \le R+2,
\]
implying that the redundancy is $\le R+2$. 
Furthermore,~\eqref{eq:wru} implies from~\cite{BGG08} that the length-$n$ redundancy of 
$\cU^\infty$ diminishes sublinearly. 
Now pick an integer $m\ge1$. We have for all $p\in\cU$,
\[
\sum_{n\in T_{p,\frac1{m^22^{m^2}}}} 
\!\!\!\!
p(n) \log \frac{p(n)}{q(n)}
\le 
\frac{R+1}{m^2},
\]
yet for all $k\ge m$, we have
\[
\sum_{n\in T_{p,\frac1{m^22^{m^2}}}} 
\!\!\!\!
p_k(n) \log \frac{1}{p_k(n)}=1.
\]
Thus it is easy to see that $\cU$ indeed satisfies~\eqref{eq:U}.

\bibliographystyle{unsrt}
\bibliography{univcod}

\begin{thebibliography}{10}

\bibitem{OSZ03}
A.~Orlitsky, N.P. Santhanam, and J.~Zhang.
\newblock Universal compression of memoryless sources over unknown alphabets.
\newblock {\em IEEE Transactions on Information Theory}, 50(7):1469---1481,
  July 2004.

\bibitem{BGG08}
S.~Boucheron, A.~Garivier, and E.~Gassiat.
\newblock Coding on countably infinite alphabets.
\newblock Available from arXiv doc id: 0801.2456, 2008.

\bibitem{SAKS14}
N.~Santhanam, V.~Anantharam, A.~Kavcic, and W.~Szpankowski.
\newblock Data driven weak universal redundancy.
\newblock Submitted for publication.
\newblock Full version available form arXiv.

\bibitem{MF98}
N.~Merhav and M.~Feder.
\newblock Universal prediction.
\newblock {\em IEEE Transactions on Information Theory}, 44(6):2124---2147,
  October 1998.

\bibitem{JR}
Jeffrey Rosenthal.
\newblock {\em A first look at rigorous probability theory}.
\newblock World Scientific, 2nd edition, 2008.

\bibitem{S06:thesis}
Narayana Santhanam.
\newblock {\em Probability estimation and compression involving large
  alphabets}.
\newblock PhD thesis, University of California, San Diego, 2006.

\bibitem{Kin80}
J.F.C. Kingman.
\newblock The mathematics of genetic diversity.
\newblock {\em SIAM}, 1980.

\bibitem{Zab92}
S.L. Zabell.
\newblock Predicting the unpredictable.
\newblock {\em Synthese}, 90:205--232, 1992.

\bibitem{Zab97}
S.L. Zabell.
\newblock The continuum of inductive methods revisited.
\newblock In John Earman and John~D. Norton, editors, {\em The Cosmos of
  Science: Essays of Exploration}, chapter~12. The University of Pittsburgh
  Press, Pittsburgh, PA, USA, 1997.

\bibitem{Zab05}
S.L. Zabell.
\newblock {\em Symmetry and Its Discontents: Essays on the History of Inductive
  Probability}.
\newblock Cambridge Studies in Probability, Induction, and Decision Theory.
  Cambridge University Press, Cambridge, 2005.

\bibitem{OS14}
A.~Orlitsky and N.~Santhanam.
\newblock Lecture notes on universal compression.
\newblock Available online from
  \texttt{http://www-ee.eng.hawaii.edu/\~{}prasadsn/}.

\end{thebibliography}
\end{document}